\documentstyle[prl,aps,psfig,graphicx,epsf,twocolumn,floats,amssymb,amsfonts]{revtex}

\def\mbf(#1){\mbox{\boldmath $#1$}}

\begin{document}

\draft

\title{Localization in ruthenates: magnetic and electronic
properties of Ca$_{2-x}$Sr$_x$RuO$_4$}

\author{~V.I.~Anisimov$^1$, ~I.A.~Nekrasov$^1$,~D.E.~Kondakov$^1$,
T.M.~Rice$^2$ and M.~Sigrist$^3$} 

\address{$^1$ Institute of Metal Physics, 
Russian Academy of Sciences-Ural Division,
620219 Yekaterinburg GSP-170, Russia \\
$^2$ Theoretische Physik , ETH-H\"onggerberg
CH-8093 Z\"urich, Switzerland \\
$^3$ Yukawa Institute for Theoretical Physics, Kyoto University,
Kyoto 606-8502, Japan}
\date{\today}
\maketitle

\begin{abstract}
\noindent
The electronic structures of the metallic and insulating phases of
Ca$_{2-x}$Sr$_x$RuO$_4$ ($0\leq x \leq 2$) are calculated using LDA,
LDA+U and Dynamical Mean-Field Approximation methods. For $x=0$ the
ground state is an orbitally non-degenerate antiferromagnetic 
insulator. For $0.2<x<0.5$ we propose a state with partial orbital and
spin ordering. For $x>0.$5 the observed Curie-Weiss paramagnetic
metallic state possessing a local moment with the unexpected spin $S=1/2$ is
explained by the localization of only two of the three Ru-4d-orbitals.
\end{abstract}

\pacs{75.30.-m,75.50.-y,71.25.Tn}

\narrowtext
  
  The isoelectronic alloy series Ca$_{2-x}$Sr$_x$RuO$_4$ is a rare
  example of a metal-insulator transition with a multi-band electronic
  structure. Sr$_2$RuO$_4$ has been well studied of late because of
  its unconventional superconductivity and strongly correlated
  Fermi-liquid behavior in the normal phase \cite{SRO,cor1,cor2}. The
  other end   member,   Ca$_2$RuO$_4$ is a localized Mott insulator,
  antiferromagnetically   (AF) ordered and with a distorted crystal
  structure \cite{maeno,CRO1,CRO2,CRO3,CRO4}.  It is very 
  interesting to examine how this isoelectronic system evolves between
  these two contrasting end members.
  
  Very recently Nakatsuji and Maeno (NM) reported transport, magnetic
  and crystallographic measurements on the intermediate phases
  \cite{maeno,CRO1}. 
  Especially noteworthy is the pronounced increase in paramagnetism in
  a metallic phase as $x$ is decreased from pure Sr$_2$RuO$_4$ $(x=2)$
  down to a critical value $x_c=0.5$.  In this range $2>x>0.5$ (called
  Region III by NM), the susceptibility evolves to show free Curie
  behavior (i.e. $x \approx C/T$) at $x=x_c$ with a Curie constant $C$
  corresponding to a spin $S=1/2$ per Ru$^{4+}$-ion.  This result
  presents two puzzles: the apparent absence of exchange interactions
  in this concentrated spin system and the unexpected reduction in the
  value of the local moment from the usual $S=1$, for localized
  Ru$^{4+}$-ions \cite{maeno}. Further reduction of $x<x_c$ leads to Region II
  characterized by a magnetic metallic phase in which both a tilting of
  the RuO$_6$ octahedra and AF correlations between the $S=1/2$
  moments appear in a metallic phase.  Finally in Region I $(0<x<0.2)$
  a metal-insulator transition occurs as the temperature is lowered.
  The end member Ca$_2$RuO$_4$ is an AF Mott insulator with the
  expected $S=1$ local moment and a more strongly distorted crystal
  structure. This complex evolution of the electronic structure from a
  well-formed Landau-Fermi liquid in Sr$_2$RuO$_4$ to the AF Mott
  insulator in Ca$_2$RuO$_4$ must be interpreted in terms of the
  multiband electronic structure of these materials.
  
  In this Letter we report calculations of the electronic structure of
  Ca$_{2-x}$Sr$_x$RuO$_4$ using local density (LDA), local density
  with onsite correlations (LDA+U) and Dynamic Mean-Field (DMFT)
  approximation schemes, in each range for compositions for which
  crystal structures are determined. In this way we track the
  evolution in the electronic structure from the metal at $x=2$ to the
  AF insulator at $x=0$.
  
  Sr$_2$RuO$_4$ has the single-layered K$_2$NiF$_4$-structure (space
  group $I4/mmm$) \cite{CRO3,Sr-struc} with layers of RuO$_6$-octahedra,
  slightly elongated along the c-axis and tetragonal local symmetry
  for a Ru$^{4+}$-ion.  The band structure has completely filled
  O-$2p$-bands, 4 electrons in Ru $t_{2g}$ 4d-band and an empty
  $e_g$-band.  Due to the tetragonal symmetry, the $xy$-orbitals of
  the $t_{2g}$-band are not equivalent to the $(xz,yz)$-orbitals. The
  crystal field splitting between them is small but the $xy$-orbitals
  $\pi$-hybridize with the $2p$-orbitals of all 4 in-plane O-neighbors
  while $xz$ ($yz$)-orbitals $\pi$-hybridize only with the 2
  O-neighbors on the x-axis ($y$-axis).  As a result the $xy$-band has
  two-dimensional dispersion and a bandwidth nearly twice that of the
  ($xz,yz$)-orbitals (see Fig.~1). The measured effective masses,
  however, are typically enhanced by factors of 3 to 4, indicating
  strong correlations. Note the volumes contained within the Fermi
  surface sheets show an approximately equal occupation of all 3
  orbitals --- a state we denote as $(8/3, 4/3)$ where $ (n_{(\alpha,
    \beta)} , n_{\gamma} )$ are the occupation of the $(yz,zx)$- and
  $xy$-orbitals, respectively.
  
  Turning to the other end member, we note that the ionic radius of
  Ca$^{2+}$ is smaller than Sr$^{2+}$, so that a stabilization of the
  metallic state might be expected. However with the substitution of
  Sr by Ca in Ca$_{2-x}$Sr$_x$RuO$_4$, the average (Sr,Ca)-ion size
  becomes too small for its space in the lattice of RuO$_6$-octahedra.
  The lattice reacts by rotating and tilting the octahedra so as to
  decrease this space. Also the smaller size of the Ca$^{2+}$-ion
  decreases the interlayer distance (c-axis lattice constant) which
  results in a smaller elongation of the RuO$_6$-octahedra (for
  Ca$_2$RuO$_4$ the octahedra are even compressed \cite{CRO3}).  These
  structural changes lead to a smaller bandwidth due to the reduction
  of the Ru-O-Ru-bond angle from the ideal value of 180$^\circ$ and
  the compression of the octahedra changes the sign of the
  $xy$-$(xz,yz)$ energy splitting.
  
  We consider the evolution of the electronic structure starting from
  the good metal, Sr$_2$RuO$_4$. Here the fractional occupation of all
  3 bands that cross the Fermi surface clearly inhibits Mott
  localization, since none of the Fermi sheets allows elastic umklapp
  electron-electron scattering at the Fermi energy. Yet the evolution
  from $x=2$ to $x=0.5$ clearly points to the partial localization of
  the 4d-electrons.  Metallic strongly correlated electron systems
  have been successfully described by the recently developed
  DMFT~\cite{pruschke,georges96}.  We have used LDA to determine the
  band parameters of the non-interacting Hamiltonian and then treated
  the interactions within the DMFT. The effective Anderson impurity
  model in the DMFT scheme has been solved by the non-crossing
  approximation (NCA)\cite{prhuschke89}.  This leads to the combined
  method called LDA+DMFT+NCA~\cite{zoelfl99}.
  
  The most puzzling aspect is the effective $S=1/2$ moment observed in
  the Curie-Weiss-like susceptibility, since the ionic value for the
  low-spin state of a Ru$^{4+}$-ion is $S=1$. In the absence of full
  crystallographic data for this region, we applied the LDA+DMFT+NCA
  scheme to the Sr$_{2}$RuO$_4$ structure, with $U$ values varying
  from 1.0 eV to 2.5 eV.  The partially filled Ru-4d-bands derived
  from the $xy$ and $(yz,zx)$ orbitals do not hybridize with each
  other by symmetry. Since the effective bandwidth of the $xy$-band is
  considerably larger (Fig.1a), a larger value of $ U $ is needed to
  drive the metal-insulator transition for the $xy$-band than for the
  $(yz,zx)$-band. The LDA+DMFT+NCA calculations showed that $U_c=1.5$
  eV is a critical value for localization of the $(yz,zx)$-electrons
  while a larger value of $U_c=2.5$ eV is needed for the $xy$-band.
  These results imply that for intermediate values of $U$ (1.5~eV $< U
  <$ 2.5~eV), there will be a regime of partial localization of the
  4d-electrons.  In practice the RuO$_6$-octahedra rotate when $x$ is
  $<1.5$, reducing the n.n. hopping and narrowing the band. For
  constant $U$, Mott localization will appear first in the
  $(xz,yz)$-band. This leads us to propose that the region $x\sim x_c$
  has a novel electronic structure illustrated in Fig.~1b. Of the 4
  electrons in the $t_{2g}$-bands, 3 electrons (or 1 hole) are in the
  Mott localized $(xz,yz)$-bands and are the origin of the $S=1/2$
  local moments. The remaining 1 electron forms a half-filled but
  still metallic $xy$-band.  Integer occupancy of the different
  sub-bands is the key to achieving Mott localization. Our LDA + DFMT
  + NCA calculations give a self-consistent solution with exactly (3,
  1) occupancy.
  
  We now turn to the magnetic properties of the local $S=1/2$ moments.
  While a Kondo-type of interaction between the two bands $(xz,yz)$
  and $xy$ can be excluded due to the absence of hybridization for
  symmetry reasons, an AF-type of RKKY interaction between the
  localized spins is still induced via Hund's rule coupling. At the
  same time, however, Hund's rule coupling causes the ferromagnetic
  correlation through the double exchange mechanism. Thus, the two
  types of spin interactions mediated by the itinerant electrons of
  the $xy$-band are likely to compensate each other, such that the
  exchange coupling between neighboring localized spins occurs mainly
  through superexchange processes in the $(yz,zx)$-band. The highly
  anisotropic hopping matrix elements between these orbitals, however,
  leads to an essential dependence of the superexchange interaction on
  the orbital configuration of the minority spin electron (or single
  hole) of each Ru$^{4+}$-ion in the degenerate $(xz,yz)$ bands. To
  gain insight into the possible form of orbital ordering we performed
  LDA+U calculations for the critical concentration $ x_c=0.5$
  \cite{Anisimov91}.

\begin{figure}[t]
\begin{center}
\begin{minipage}[t]{8cm}
\epsfxsize=8cm
\epsfbox{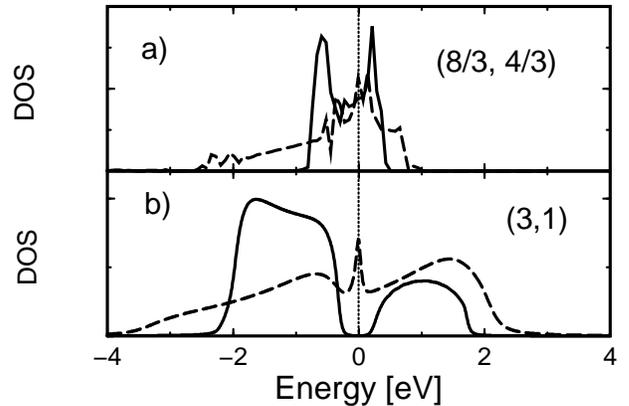}
%Fig.~1

\caption {Density of states (DOS) for Sr$_2$RuO$_4$: a)
The partial DOS obtained from the LDA calculation. b)
The results of LDA+DMFT+NCA calculations for paramagnetic
metallic state. The parameters for non interacting Hamiltonian
were calculated using Sr$_2$RuO$_4$ crystal structure and
Coulomb interaction parameter $U$=2 eV.
The solid line is the DOS for $(xz,yz)$-orbitals and dashed line
for $xy$-orbital. $ (n_{(yz,zx)}, n_{xy} ) $ indicates the electron
distributions. }
\end{minipage}
\end{center}
\label{dos} 
\end{figure}

Ca$_{1.5}$Sr$_{0.5}$RuO$_4$ has the space group $I4_1/acd$
\cite{Ca-struc} which has no tilting of RuO$_6$-octahedra, only
rotations around the $c$-axis.  The RuO$_6$-octahedra are also
elongated causing the $xy$-orbital to be higher in energy than
$(xz,yz)$-orbitals leading to orbital degeneracy in a (3, 1) state.
The LDA+U calculation gave an orbital order of an antiferro-type. The
minority-spin electrons occupy alternating $xz$- and $yz$-orbitals
$(\langle\tau^z\rangle =\pm 1)$ with a slight tilting of the orbital
planes away from the c-axis. This orbital order favors ferromagnetic
(FM) spin alignment and this could explain the experimentally observed
enhancement of susceptibility with increasing of Sr concentration
toward the $x=0.5$ \cite{maeno}.

Specifically the localized orbital degrees of freedom are represented
by the isospin $ \mbox{\boldmath $ \tau $} $ where the $ yz $- and $
zx $-orbitals on each site correspond to the states $ | + \rangle $
and $ | - \rangle $, respectively, with $ \tau^z | \pm \rangle = \pm
\frac{1}{2} | \pm \rangle $. We parameterize the n.n.  superexchange
interaction as
\begin{equation}
{\cal H}_s = \sum_{i, {\bf a}}  J_{{\bf a}}( \mbox{\boldmath $ \tau 
$}_i , \mbox{\boldmath $ \tau 
$}_{i+ {\bf a}} ) \{ {\bf s}_i \cdot {\bf s}_{i + {\bf a}} +b \}
\label{heis}
\end{equation}
with $ J_{{\bf a}}( \mbox{\boldmath $ \tau $}_i , \mbox{\boldmath $
  \tau $}_{i+ {\bf a}} ) = J_1 {\bf I}_{i,{\bf a}} \cdot {\bf
  I}_{i+{\bf a},{\bf a}} $. We define the orbital (2-component) vector
operator $ {\bf I}_{i, {\bf a}} = (\tau^z_i + g \alpha \tau^0_i,
\tau^x_i + g \alpha' \tau^y_i ) $ where $ {\bf a} = (1,0),(0,1)$ is
the vector connecting n.n. sites and $ g = a_x^2 - a_y^2 $ ($ \tau^0 =
\frac{1}{2} \hat{1}_{2 \times 2} $). This Hamiltonian is reduced to
the essential parts from the most general possible form. The
parameters $ \alpha $, $ \alpha' $, $ b $ and $ J_1 $ are chosen to be
consistent with the results of the LDA calculation. In particular, $
0< 1-\alpha \ll 1 $ arising from the reduction of the overlap for
orbitals with their planes $\perp \vec{\bf a}$. This leads
to AF coupling, if the orbital states on a bond are the same (strongly 
reduced, if the bond is perpendicular to the plane), and to FM coupling
for the configuration $ |+\rangle \otimes | - \rangle $.  This
competition between AF and F coupling suggests that there may be a
strong cancellation due to fluctuations of the orbital occupation.

At lower values of $x$ we enter Region II $(0.5>x>0.2)$ characterized
by a tilting plus rotation of RuO$_6$-octahedra.
Ca$_{1.8}$Sr$_{0.2}$RuO$_4$ has a low-symmetry crystal structure with
the space group $P2_1/c$ \cite{Ca-struc}, which can be obtained from
the tetragonal $I4/mmm$ structure by rotating and tilting of the
RuO$_6$-octahedra similar to pure Ca$_2$RuO$_4$ but with a smaller
tilting angle \cite{Ca-struc}.  There are now two types of in-plane
oxygen ions and two types inequivalent of RuO$_6$-octahedra. The ratio
of the Ru-O bond length for the apical and planar oxygens is larger
than 1, so that the $xy$-level lies higher than the $(xz,yz)$-levels
leading to an occupation (3, 1).  The single minority-spin electron is
orbitally degenerate in $(yz,zx)$-band.  We performed LDA+U
calculation for this tilted structure and obtained a rather
complicated orbital order. The ground state is an AF insulator.  The
minority-spin electron (1 per Ru-atom) occupy the orbitals whose
planes are in average directed along the $a$-axis (in tetragonal
notation (110) direction).  However on every one of the 4 Ru-atoms in
the unit cell those planes are rotated from the $a$-axis by
+20$^\circ$ ($ \langle \mbox{\boldmath $ \tau $} \rangle = (u,0,v) $)
and +15$^\circ$ ($ \langle \mbox{\boldmath $ \tau $} \rangle =
(u',0,v') $) on one layer and by -20$^\circ$ ($ \langle
\mbox{\boldmath $ \tau $} \rangle = (-u,0,v) $) and -15$^\circ$ ($
\langle \mbox{\boldmath $ \tau $} \rangle = (-u',0,v') $) on the next
layer ($ u = 0.643 $, $ v = 0.766 $, $ u' = \sqrt{3}/2 $ and $ v'=1/2
$). Also on one of the 2 Ru-atoms in every layer there is an
additional tilting of the orbital plane from the long $c$-axis on
34$^\circ$ which is not represented within our model. The calculation
of the easy axis using the second order perturbation theory for
spin-orbit coupling gave the direction of magnetic moment as along the
$a$-axis (tetragonal $ [1\bar{1}0] $ direction) with a 28$^\circ$ tilt
from the layer plane. The measurements of the uniform magnetic
susceptibility show a peak in the temperature dependence which is the
most pronounced for the $ [110]$ direction of the magnetic field, in
agreement with our LDA+U results \cite{maeno}.

To include the aspect of spin anisotropy we extend the
interaction adding the following terms,

\begin{equation} \begin{array}{l}
\sum_{i,{\bf a}} [J_2 g (I^x_{i, {\bf a}} I^x_{i+{\bf
a},{\bf a}} - I^y_{i, {\bf a}} I^y_{i+{\bf a},{\bf a}})(s^x_i s^x_{i
+{\bf a}} - s^y_i s^y_{i+{\bf a}}) \\ \\
\quad + J_3 g ((I^x_{i, {\bf a}} I^y_{i+{\bf
a},{\bf a}} + I^y_{i, {\bf a}} I^x_{i+{\bf a},{\bf a}})(s^x_i s^y_{i
+{\bf a}} + s^y_i s^x_{i+{\bf a}}) \\ \\
\quad + J_4 ((I^x_{i, {\bf a}} I^y_{i+{\bf
a},{\bf a}} - I^y_{i, {\bf a}} I^x_{i+{\bf a},{\bf a}})(s^x_i s^y_{i
+{\bf a}} - s^y_i s^x_{i+{\bf a}}) \\ \\
\quad +J_5 s^z_i s^z_{i+{\bf a}} ] 
\label{anis} \end{array} \end{equation}
where further phenomenological parameters appear.
Our previous discussion based on LDA indicated
that crystal field effects yield the dominant bias for the orbital
orientation. This is taken into account by an additional term $ {\cal
H}_{cf} = -  \sum_{i, \mu} Q^{\mu}_i \tau^{\mu}_i $ where $ {\bf Q}_i
$ is a local``Jahn-Teller'' mean field which depends on the
distortion introduced by Ca-doping and may 
show a complicated staggering as suggested by our LDA calculation.

We now can connect the experimental results for $ x = 0.2 $ with our
model. Taking for simplicity $ {\bf Q}_i \propto (1,0,0)$ uniformly
plane of all orbitals is parallel to $ [1 \bar{1} 0] $ and we obtain
an effective AF anisotropic n.n. spin exchange of the form $
\sum_{i,{\bf a},\mu} \tilde{J}_{\mu} s^{\mu}_i s^{\mu}_{i +{\bf a}} $.
Here the coordinates $ \mu $ are the principal spin axis in the
exchange (1) and (2). Two axis ($ \mu = x'$ and $ z $) lie in the
plane of the orbitals and the third axis ($ \mu = y' $) lies
orthogonal. Assuming $ J_3 > 0 $ and $ J_5 \approx J_3 $ leads to an
``XXZ-type'' Heisenberg model where the Ising part (spin quantization
axis along $ [1 1 0] $) is weaker. Consequently, the uniform
susceptibility is larger along the $ [1 1 0] $-direction. The
temperature dependence of the susceptibility for all directions show
peak features indicating the characteristic energy scale which is
lower for [110] than the other two, consistent with the different AF
exchange in our model.  While the shape of the susceptibility for $
[110] $ and $ [001] $ are very similar the latter direction is
enhanced by a constant. This can be attributed to the
temperature-independent orbital Van Vleck contribution along the $
z$-axis, related to the orbital operator $ \tau^y $.

Lastly we consider the end member Ca$_2$RuO$_4$.  The structure of
Ca$_2$RuO$_4$ is orthorhombic (space group $Pbca$) \cite{CRO3} and is
obtained from the high-symmetry tetragonal $I4/mmm$ by rotating the
RuO$_6$-octahedra around their long axis (001) and tilting around the
diagonal in-plane axis (110), with the condition that all planar
oxygens move the same distance from the plane. As a result there is
only one type of in-plane oxygen in this structure and all
RuO$_6$-octahedra are equivalent.

Ca$_2$RuO$_4$ is an AF insulator and is relatively simple to describe
with the LDA+U method \cite{Anisimov91}. This method is based on
spin-orbital unrestricted Hartree-Fock equations (i.e. a {\it static}
mean-field approximation) and generally gives good results for long
range ordered Mott insulators \cite{Anisimov97}. The Coulomb direct
and exchange interaction parameters values used in our calculation
were $U$=1.5 eV and $J$=0.7 eV. The groundstate of Ca$_2$RuO$_4$ was
found to be an AF insulator with a sublattice moment of 1.51 $\mu_B$
and a small energy gap, 0.30 eV, which compares well with the
experimental values 1.3 $\mu_B$ \cite{CRO3} and 0.2 eV \cite{CRO4}
respectively.  The band $t_{2g}$ is now fully spin-polarized (the
reduced moment relative to $S=1$ is due to hybridization with oxygen
states). Of the 4 $t_{2g}$-electrons, 3 are in the majority-spin band
and 1 is in the minority-spin band. In general, this implies an
orbitally degenerate state.  However in Ca$_2$RuO$_4$, this degeneracy
is lifted, since the $xy$-level lies considerably lower than the
($xz,yz$)-levels and is fully occupied in the insulating state.  In
the notation introduced above, we have integer (2, 2) occupancy of the
orbitals in contrast to the (3, 1) occupancy found for Regions III and
II with $x > 0.2$. Using second order perturbation theory for
spin-orbit coupling, we find the magnetic moment along the $b$-axis in
orthorhombic notations (or [110] direction in a tetragonal notation),
in agreement with the experimental data \cite{CRO3}. The AF insulating phase
extends to $ x =0.2 $ where by a first order structural change occurs
to Region II ($0.2 \le x < 0.5 $) which exhibits short-range AF
correlation and is metallic \cite{maeno}. This first order boundary marks the
transition from the (2, 2) to (3, 1) configurations with increasing
$x$.
\begin{table}
\begin{tabular}{l|cccc}
Region & ($ n_{yz,zx} , n_{xy} $) & orbital & spin & order  \\
\hline
I $ (0 \leq x < 0.2) $ & (2,2) & -&  $ S=1$ & AF \\
II $ (0.2 \leq x < 0.5 ) $& (3,1) & $ (yz,zx) $ & $ S= \frac{1}{2} $ &
FO, (AF) \\ 
III $(x \to 0.5)$ & $ (3,1) $ & $ (yz,zx) $ &  $ S
\to \frac{1}{2} $ &  (AFO) , (FM) \\
III $ (x=2) $ & $ \left(\frac{8}{3}, \frac{4}{3} \right) $ & - & S=0 & - 
\\
\end{tabular}
\caption{The three Regions of Ca$_{2-x}$Sr$_x$RuO$_4$ with their
orbital occupancy, the localized orbital and spin degrees of freedom
and the order or (dominant correlations) (AF = spin antiferromagnetic,
FM = spin ferromagnetic, AFO = antiferro-orbital, FO = ferro-orbital).}
\end{table}
In summary, we have presented a consistent picture for the unusual
phases of the isoelectronic alloy series Ca$_{2-x}$Sr$_2$RuO$_4$ based
on the full multi-band electronic structure (see Table I). Starting
from the good 
metal Sr$_2$RuO$_4$, we find the effect of Ca-substitution is to
transfer electrons from the wider $(xy)$-band to the narrower
$(xz,yz)$ bands until at a critical value of $x_c=0.5$ there is
integer occupancy of both subbands. The progressive rotation of the
RuO$_6$ octahedra in this region leads to Mott localization of the 3
electrons in the narrower $(xz,yz)$ bands while the wider $xy$-band
which is now half-filled, remains metallic. This partial localization
of the 4d electrons can explain the puzzling observation of the
coexistence of free $S=1/2$ local moments and metallic behavior in
Ca$_{1.5}$Sr$_{0.5}$RuO$_4$.  The ordered phases appearing in the
Ca-rich Regions I ($ 0 \leq x < 0.2 $) and II ($ 0.2 \leq x <0.5 $),
are based on localized ($ yz,zx$)-orbitals, while the $ xy $-orbital
plays the role of a charge reservoir, yielding a completely filled
(insulating)or a half-filled (metallic) band in the two regions,
respectively. In the former case the ordered local spin has $S=1$, but
in the latter case short-range correlations of a local $S=1/2$ spin
combined with orbital order are realized, generating a pronounced
anisotropy in the magnetic response.

\bigskip

We would like to thank Y. Maeno, S. Nakatsuji, T. Nomura , M. Braden
and A. Poteryaev for many helpful discussions.  This work was
financially supported by the Russian Foundation for Basis Research
grant RFFI-98-02-17275 and a Grant-in-Aid of the Ministry of
Education, Science, Sports and Culture of Japan.


\begin{references}


\bibitem{SRO}Y. Maeno , H. Hashimoto, K.Yoshida, S. Nishizaki,
T. Fujita, J.G. Bednorz, F. Lichtenberg, Nature (London) {\bf 372},
532 (1994)
\bibitem{cor1} A.P. Mackenzie, S.R. Julian, A.J. Diver, G.J. McMullan,
M.P. Ray, G.G. Lonzarich, Y. Maeno, S. Nishizaki, T. Fujita,
Phys. Rev. Lett. {\bf 76}, 3786 (1996)
\bibitem{cor2} Y. Maeno, K. Yoshida, H. Hashimoto, 
S. Nishizaki, S. Ikeada, M. Nohara, T. Fujita, A.P. Mackenzie,
N.E. Hussey, J.G. Bednorz, F. Lichtenberg, J. Phys. Soc. Jpn. {\bf 66},
1405 (1997)
%\bibitem{Ir} R.J. Cava, B. Batlogg, K. Kiyono, H. Takagi,
%J.J. Krajewski, W.F. Peck, Jr., L.W. Rupp, Jr., C.H. Chen,
%Phys. Rev. B {\bf 49}, 11 890 (1994)
%\bibitem{Ir2} M.K. Crawford, M.A. Subramanian, R.L. Harlow, J.A. 
%Fernandez-Baca, Z.R. Wang, D.C. Johnston, Phys. Rev. B {\bf 49}, 9198 (1994);
\bibitem{maeno} S. Nakatsuji, Y. Maeno, Phys. Rev. Lett. {\bf 84},
2666 (2000).
\bibitem{CRO1}S. Nakatsuji, S. Ikeda, Y. Maeno, J. Phys. Soc. Jpn. {\bf 66},
1868 (1997)
\bibitem{CRO2} G. Cao, S. McCall, M. Shepard, J.E. Crow, R.P.Guertin,
Phys. Rev. B {\bf 56}, R2916 (1997)
\bibitem{CRO3} M. Braden, G. Andre, S. Nakatsuji, Y. Maeno,
Phys. Rev. B {\bf 58}, 847 (1998)
\bibitem{CRO4} A.V. Puchkov, M.C. Schabel, D.N. Basov, T. Startseva,
G. Gao, T. Timusk, Z.-X. Shen, Phys. Rev. Lett. {\bf 81}, 2747 (1998)
\bibitem{Sr-struc} M. Braden, H. Moudden, S. Nishizaki, Y. Maeno,
T. Fujita, Physica C {\bf 273}, 248 (1997)
\bibitem{Ca-struc} O. Friedt, M. Braden, G. Andr\'e., P. Adelmann,
S. Nakatsuji and Y. Maeno, in preparation. 
\bibitem{pruschke}Th.\ Pruschke, M.\ Jarrell and J.\ K.\ Freericks,
  Adv. in Phys. {\bf 44}, 187 (1995)
\bibitem{georges96} A. Georges, G. Kotliar, W. Krauth
and M.J. Rozenberg, Rev. Mod. Phys. {\bf 68}, 13 (1996).
\bibitem{prhuschke89} Th. Pruschke and N. Greve, Z. Phys. B {\bf 74},
439 (1989).
\bibitem{zoelfl99} M.B. Z\"olfl, Th. Pruschke, J. Keller,
A.I. Poteryaev, I.A. Nekrasov and V.I. Anisimov, Phys. Rev. B {\bf 61}, 
12810 (2000).
\bibitem{Anisimov91} V.I. Anisimov, J. Zaanen and O.K. Andersen,
Phys. Rev. B {\bf 44}, 943 (1991)
\bibitem{Anisimov97} V.I. Anisimov, F. Aryasetiawan and 
A.I. Lichtenstein,
J. Phys.: Condens. Matter {\bf 9}, 767 (1997).


\end{references}
\end{document}